\documentclass[aps,prb, onecolumn,nofootinbib,citeautoscript,10pt]{revtex4-2}  
\usepackage{dcolumn}
\usepackage{bm}
\usepackage{amsmath}
\usepackage{tabularx, graphicx}
\usepackage{epstopdf}
\usepackage{latexsym}
\usepackage{amssymb}
\usepackage{amsmath}
\usepackage{color, colortbl}
\usepackage{psfrag}
\usepackage{bbm}
\usepackage{titlesec}
\usepackage{dsfont}
\usepackage{slashed}
\usepackage{multirow}
\usepackage[tight]{subfigure}

\usepackage[papersize={8.5in,11in}]{geometry}

\usepackage{color}
\definecolor{darkblue}{rgb}{0.,0.,0.4}
\definecolor{darkred}{rgb}{0.5,0.,0.}
\definecolor{BlueViolet}{RGB}{138,43,226}
\definecolor{SkyBlue}{RGB}{30,144,255}
\definecolor{DarkGreen}{RGB}{0,100,0}
\usepackage[pdftex,colorlinks=true,linkcolor=darkblue,citecolor=blue,urlcolor=darkred]{hyperref}

\usepackage{physics}

\def \nn{\nonumber \\}

\begin{document}

\title{Linear response of tilted anisotropic two-dimensional Dirac cones}

\author{Ipsita Mandal}
\email{E-mail address: ipsita.mandal@gmail.com}
\affiliation{Department of Physics, Shiv Nadar Institution of Eminence (SNIoE),
Gautam Buddha Nagar, Uttar Pradesh 201314, India}

\begin{abstract}
We investigate the behaviour of the linear-response coefficients, when in-plane electric field ($\mathbf E$) or/and temperature gradient ($\nabla_{\mathbf r} T$) is/are applied on a two-dimensional semimetal harbouring anisotropic Dirac cones. The anisotropy is caused by (1) differing Fermi velocities along the two mutually perpendicular momentum axes, and (2) tilting parameters. Using the semiclassical Boltzmann formalism, we derive the forms of the response coefficients, in the absence and presence of a nonquantizing magnetic field $\mathbf B$. The magnetic field affects the response only when it is oriented perpendicular to the plane of the material, with the resulting expressions computed with the help of the so-called Lorentz-force operator, appearing in the linearized Boltzmann equation. The solution has to be found in a recursive manner, which produces terms in powers of $|\mathbf B|$. We discuss the validity of the Mott relation and the Wiedemann-Franz law for the Lorentz-operator-induced parts.
\end{abstract}

\maketitle

\tableofcontents


\section{Introduction}

The establishment of graphene as a poster-child for two-dimensional (2d) semimetals \cite{Castro2009} is mainly due to the appearance of Dirac cones (featuring linear-in-momentum dispersion), at half-filling of the underlying honeycomb lattice. Since the discovery of graphene, emergence of novel features in experimental and theoretical investigations of graphene has never ceased to intrigue us. This has been accompanied by the search and discovery of Dirac cones in other 2d materials like the organic compound $\alpha$-(BEDT-TTF)$_2$I$_3$ \cite{org1, org2}, under the action of uniaxial strain \cite{org3} or hydrostatic pressure \cite{org4}. While the Dirac cones of graphene are isotropic near half-filling, those in $\alpha$-(BEDT-TTF)$_2$I$_3$ are anisotropic for two reasons: they are tilted and they have direction-dependent Fermi velocities \cite{goerbig, thesis-fuchs}, as shown schematically in Fig.~\ref{figdis}. A quinoid-type lattice distortion of a graphene sheet \cite{pauling} or T-graphene under uniaxial strain \cite{hancock} also exhibits such distortion in the emergent Dirac cones. Furthermore, anisotropic Dirac cones can be engineered in ultracold atoms in a hexagonal optical lattice, with the lattice distortion, relevant hopping parameters, etc. being controlled via laser intensities, wavelengths, and orientation-adjustments \cite{op-lat}. While anisotropic nearest-neighbor hoppings in a tight-binding model on the honeycomb lattice cause the Dirac points to move away from the high-symmetry points (i.e., the K and K$^\prime$ corners) towards one of the three M points, tilting of the Dirac cones occurs when diagonal terms are present in the basis formed by the A and B sublattice points (e.g., next-nearest neighbour hoppings \cite{goerbig}). On increasing the applied strain, at some critical value, the two valley-conjugate Dirac points eventually merge, and the low-energy spectrum shows a hybrid dispersion \cite{montambaux, ips-kush, thesis-fuchs, ips-kush-review}. The mixed nature is featured by a semi-Dirac semimetal, whose dispersion is linear and quadratic along the two mutually perpendicular directions, demonstrating novel features in transport \cite{ips-kush, ips-kush-review, ips-cd}.

The anisotropic dispersion property described above allows us to play around with the direction-dependence of the transport properties, when a sample is subjected to probe fields like electric field ($\mathbf E $) or/and temperature gradient ($\nabla_{\mathbf r} T $). In addition, an externally applied magnetic field ($\mathbf B $) will give rise to magnetoelectric or/and magnetothermal effects \cite{ips-kush, ips-kush-review}. In this paper, we focus on the linear response originating from such configurations, considering the cases of either zero or a weak nonquantizing magnetic field.\footnote{See Refs.~\cite{ips-rahul-tilt, ips-tilted} for an analogous discussion of direction-dependent transport for the three-dimensional (3d) versions comprising tilted Weyl cones.} The qualification of ``weak'' magnetic field implies that we assume the magnitude of $\mathbf B$ to be small, leading to a small cyclotron frequency $\omega_c=e\,B/(m^*\, c) $ (where $B = |\mathbf B|$, $m^* $ is equal to the effective mass of the itinerant quasiparticles or dressed electrons). Precisely, the essential condition to ensure that we can ignore the consideration of quantized Landau levels, is given by $\hbar \, \omega_c \ll |\mu| $, where $\mu$ is the Fermi level [i.e., the energy at which the chemical potential cuts the Dirac cone(s)]. Here, we will apply the semiclassical Boltzmann formalism and the relaxation-time approximation in order to derive the form of the response tensors.

\begin{figure*}[t]
\centering
\includegraphics[width=0.65 \textwidth]{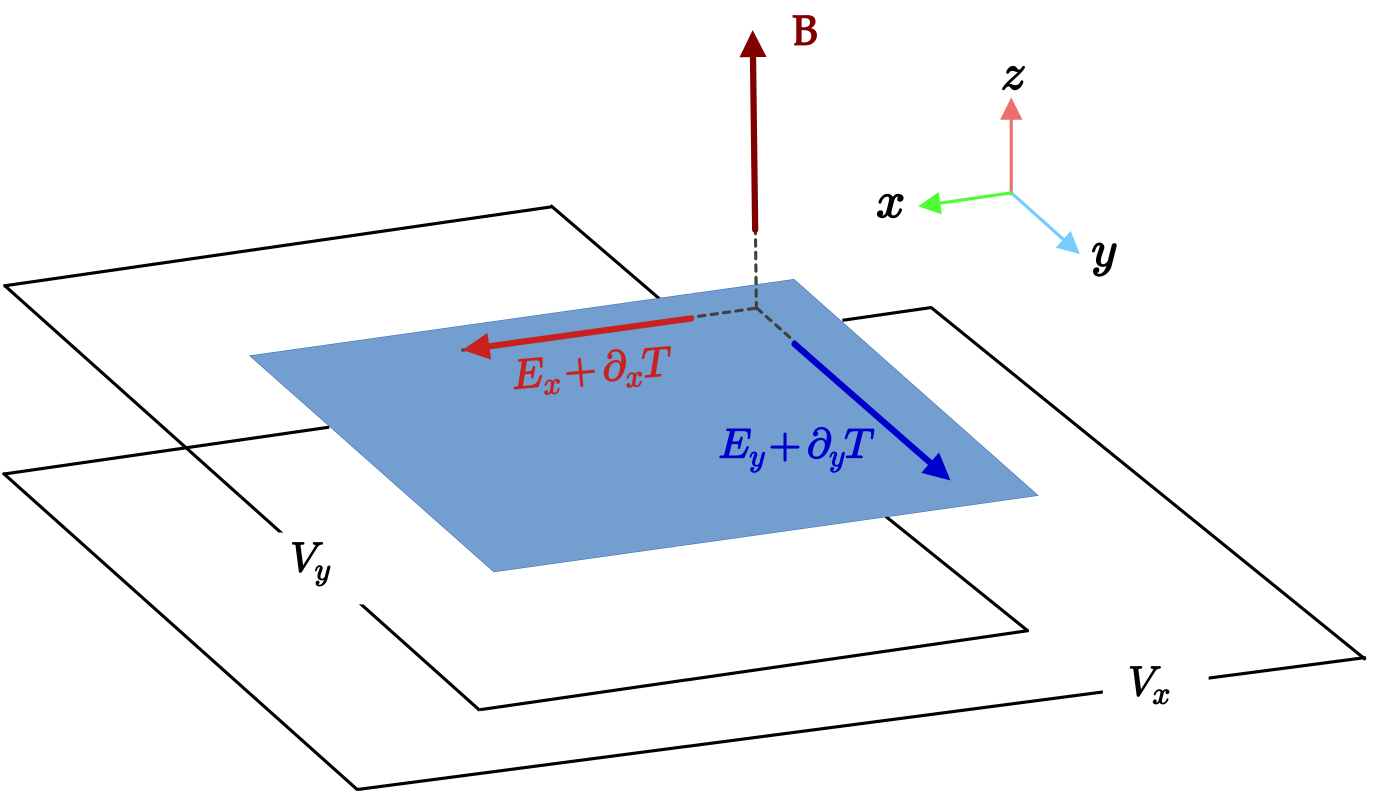}
\caption{Schematics of the setup showing the applied electric field ($\mathbf E$), temperature gradient ($\nabla_{\mathbf r} T$), and the out-of-plane magnetic field ($\mathbf B$). $V_x$ and $V_y$ represent the in-plane voltages.
\label{figsetup}}
\end{figure*}

The paper is organized as follows: In Sec.~\ref{secmodel}, we describe the effective continuum Hamiltonian in the vicinity of an anisotropic Dirac cone in 2d. We also review the generic setup of the Boltzmann equations, and show the structures of the electric and thermal currents. While Sec.~\ref{seczerob} deals with deriving the response coefficient for $\mathbf B  =\mathbf 0$, Sec.~\ref{secb} is devoted to investigating the nonzero tensors for a nonzero nonquantizing magnetic field being present [cf. Fig.~\ref{figsetup}]. We end with a summary and outlook in Sec.~\ref{secsum}, and explain some details of the intermediate steps via the appendices. In all our discussions, we will resport to using the natural units, which implies that we set the reduced Planck's constant ($\hbar $), speed of light ($c$), and the Boltzmann constant ($k_B $) to unity.

\section{Model}
\label{secmodel}

\begin{figure*}[t]
\centering
\subfigure []{\includegraphics[width=0.25 \textwidth ]{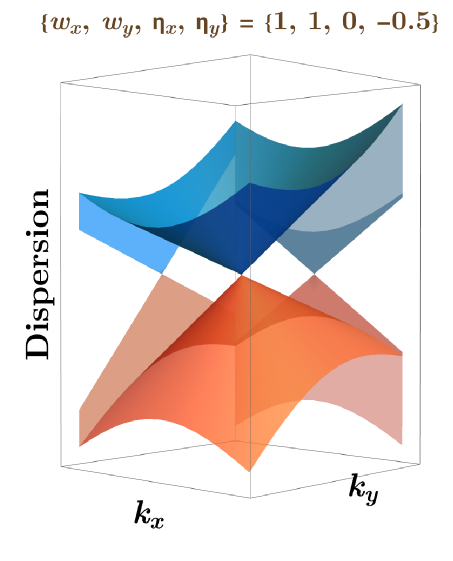}} \qquad
\subfigure []{\includegraphics[width=0.25 \textwidth ]{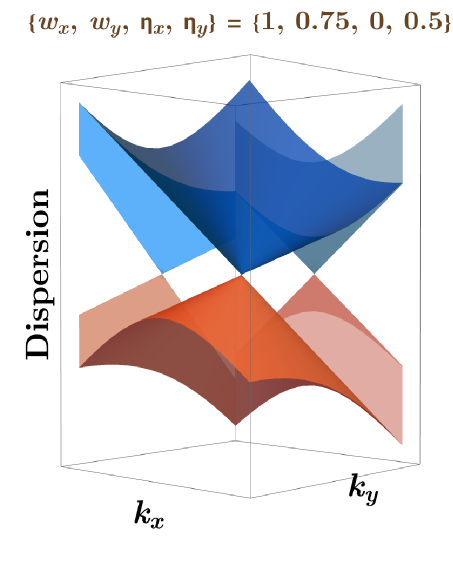}} \qquad
\subfigure []{\includegraphics[width=0.25 \textwidth ]{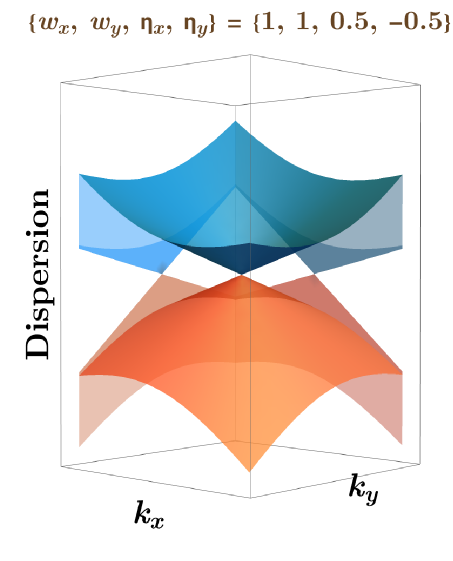}}
\caption{Schematic dispersion of quinoid graphene for some representative values of the parameters in the Hamiltonian [cf. Eq.~\eqref{eqham}], as shown in the plot-labels. The direction-dependent features are depicted with greater clarity through the the projections of the dispersion
along the respective momentum axes.
\label{figdis}}
\end{figure*}

The minimal low-energy effective Hamiltonian of quinoid-type distorted graphene sheet is captured by~\cite{goerbig}
\begin{align}
\label{eqham}
\mathcal{H} = \left ( \boldsymbol{w}_0 \cdot\mathbf{k} \right ) \sigma_0
+   w_x\, k_x \,\sigma_x +    w_y \, k_y \,\sigma_y\,, \quad
\boldsymbol{w}_0 \equiv  
w_{0x} \,   {\boldsymbol{\hat e}}_x +  w_{0y} \,  {\boldsymbol{\hat e}}_y\,, \quad
w_{0x} \equiv   w_x\,\eta_x \, ,\quad
w_{0y} \equiv  w_y\,\eta_y \,,
\end{align}
where $\sigma_0$ is the $2\times2$ identity matrix, $\sigma_x $ and $\sigma_y $ denote the first two Pauli matrices, 
and $ \boldsymbol{w}_0  $ is the vector whose two components parametrize the amount of tilt with respect to the two momentum axes, $k_x$ and $k_y$. Furthermore, $w_x$ and $w_y$ represent the components of the Fermi velocities along the two axes, and ${\boldsymbol{\hat e}}_i$ represent the unit vector along the $i^{\rm th}$ direction of a 3d Cartesian coordinate system. Here, we focus on the tilt-parameter regimes satisfying $ \eta <1$, where $\eta \equiv \sqrt{ {\eta}_x^2 + {\eta}_y^2  }$, such that we are looking at the type-I (or undertilted) cases, when the Fermi surfaces at finite values of the chemical potential yield ellipses.
The energy eigenvalues of the two bands are obtained as
\begin{align}
\epsilon^\lambda ({\mathbf k}) = \boldsymbol{w}_0 \cdot \mathbf{k} 
+ \lambda \, \varepsilon \,, \text{ where } 
\varepsilon = \sqrt{ w^2_x \, k^2_x + w^2_y \, k^2_y }
\text{ and } \lambda=\pm\,.
\end{align}
This dispersion relation is illustrated schematically in Fig.~\ref{figdis}.
The corresponding group velocity is captured by
\begin{align}
\boldsymbol{v}^\lambda ({\mathbf k}) \equiv
\nabla_{\mathbf k} \epsilon^\lambda(\mathbf{k})
=  \boldsymbol{w}_{0} + \frac{ \lambda  } {\varepsilon} 
\left(
w_x^2 \, k_x \,  {\boldsymbol{\hat e}}_x
+ w_y^2 \, k_y \,  {\boldsymbol{\hat e}}_y \right).
\end{align}
We will be using the symbol $ f_0 (\epsilon) $ to denote the Fermi-Dirac distribution function
$ \left( 1+e^{\frac{\epsilon-\mu} { T }  } \right)^{-1} $ at chemical potential $\mu$ and temperature $ T$, where have suppressed the $\mu$ and $T$ dependence for uncluttering of the notations. Furthermore, we will use a ``prime'' superscript to denote differentiation with respect to the variable shown within the brackets [for example, $ f_0^{\prime} (\epsilon^\lambda (\mathbf k))   \equiv \partial_\epsilon f_0 (\epsilon)$].

Let the average DC electric- and thermal-current densities from the quasiparticles be ${\mathbf J}$ and ${\mathbf J}^{{\rm th} } $, respectively. The linear-response matrix, which relates the resulting generalized current densities to the driving electric potential gradient and temperature gradient, is expressed as
\begin{align}
\label{eqcur1}
\begin{bmatrix}
J_i \vspace{0.2 cm} \\
{J}_i^{{\rm th}}
\end{bmatrix} & = \sum \limits_j
\begin{bmatrix}
 \sigma_{ij} &   \alpha_{ij} 
\vspace{0.2 cm}  \\
T  \,  \alpha_{ij}  &   \ell_{ij} 
\end{bmatrix}
\begin{bmatrix}
E_j
\vspace{0.2 cm}  \\
- \, { \partial_{j} T } 
\end{bmatrix} ,
\end{align}
where $ \lbrace i, j \rbrace  \in \lbrace x,\, y \rbrace $ indicates the Cartesian components of the current-density vectors and the response tensors in 2d. The notations $\sigma  $ and $\alpha $ represent the electric conductivity and the thermoelectric conductivity, respectively. The latter determines the Peltier ($\Pi  $), Seebeck ($ S $), and Nernst coefficients. The third tensor, $\ell $, represents the linear response relating the thermal-current density to the temperature gradient, at a vanishing electric field. $ S  $ , $\Pi $, and the thermal coefficient $\kappa  $ (which provides the coefficients between the heat-current density and the temperature gradient at vanishing electric current) are related as \cite{mermin, arovas, ips-kush-review}:
\begin{align}
\label{eq:kappa}
S_{ij}  = \sum \limits_{ i^\prime} \sigma  ^{-1}_{ i i^\prime }
\alpha_{ i^\prime j} \, , \quad
\Pi_ {ij} = T \sum \limits_{ i^\prime}
\alpha_{ i   i^\prime}   
\, \sigma^{-1}_{ i^\prime j} \,,\quad 
\kappa_ {ij} = \ell_{ ij }
- T \sum \limits_{ i^\prime, \, j^\prime }
 \alpha_{ i  i^\prime } \, \sigma^{-1}_{  i^\prime  j^\prime }
\, \alpha_{ j^\prime j}  \,.
\end{align}
Overall, we have three independent linear-response tensors to be computed.

We use the semiclassical Boltzmann formalism \cite{mermin, arovas, ips-kush-review, ips-rahul-ph, ips-rsw-ph} to determine the transport coefficients shown above. Let us define the nonequilibrium distribution function for the fermionic quasiparticles occupying a Bloch band labelled by the index $\lambda$, with the crystal momentum $\mathbf k$ and dispersion $\epsilon^\lambda (\mathbf k)$, by $ f^\lambda ( \mathbf r , \mathbf k, t) $.
For static and spatially-uniform probe fields in the form of $\mathbf E $ and $\nabla_{\mathbf r} T $, in the presence of a time-independent and uniform nonquantizing magnetic field $\mathbf B $, essentially the functional dependence of $  f^\lambda ( \mathbf r , \mathbf k, t)$ reduces to $  f^\lambda (\mathbf k) $. Therefore, the \textit{linearized Boltzmann equation} is given by
\begin{align}
\label{eqbol}
& \left [
{\boldsymbol{v}}^\lambda (\mathbf k)
\cdot \left \lbrace  
e \, \mathbf E + \frac{  \epsilon^\lambda (\mathbf k) - \mu } {T}  \, \nabla_{\mathbf r} T 
\right \rbrace \right]
\left [- f_0^\prime(\epsilon^\lambda (\mathbf k)) \right ]
- e \, \check{L}(\mathbf k) \, \delta f^\lambda (\mathbf k)
 = \frac{  -\,  \delta f^\lambda (\mathbf k)} 
{\tau   }  \,, \nn & \text{where }
f^\lambda(\mathbf r,\mathbf k, t) 
	=  f_0 (\epsilon^\lambda (\mathbf k) ) +  \delta  f^\lambda (\mathbf r,\mathbf k)
\text{ and } \check{L}  (\mathbf k)
\equiv  \left [ {\boldsymbol{v}}^\lambda (\mathbf k) \cross {\mathbf B}
 \right ] \cdot \nabla_{\mathbf k}\,.
\end{align}
Here, $\check{L}$ is the so-called Lorentz-force operator, which gives rise to terms corresponding to the classical Hall effects.
The linearization implies linear-order in the smallness parameter, which is proportional to the magnitudes of $\mathbf E $, $\nabla_{\mathbf r } T$, and the resulting $\delta  f^\lambda (\mathbf r,\mathbf k)$ [parametrizing the deviation of the fermionic distribution function from the equilibrium value of $f_0 (\epsilon^\lambda (\mathbf k)) $].
The above expression is valid for a relaxation-time approximation for the collision integral, which involves using a momentum-independent relaxation time $\tau$. This implies that we will treat $\tau $ as a phenomenological parameter, determined by the dominant processes of collisions.

Let us parametrize the deviation in $ f^\lambda (\mathbf{k})$ as
\begin{align}
\delta f^\lambda (\mathbf{k}) 
 = \left [- f_{0}^\prime (\epsilon^\lambda (\mathbf k)) \right ] {\tilde g}^{\lambda}  (\mathbf{k})\, .
\end{align} 
Then, Eq.~\eqref{eqbol} leads to \cite{ips-rsw-ph, ips-spin1-ph}
\begin{align}
\label{eqbolintra}
\frac{{\tilde g}^{\lambda} (\mathbf{k}) } 
{  \tau }
= -\, \sum_{n = 0}^{\infty}
\left[ e \, \tau \hat{L} (\mathbf k) \right ]^n 
\left [   \boldsymbol{v}^\lambda (\mathbf k) \cdot 
\left  \lbrace e\, \mathbf{E} +
\frac{  \epsilon^\lambda (\mathbf k)  - \mu } {T}  
\, \nabla_{\mathbf r} T  \right \rbrace  \right ],
\end{align}
which we solve for ${\tilde g}^\lambda (\mathbf{k})$ recursively. While the magnetic-field-independent part of the linear response is obtained by picking up the $n=0$ term on the right-hand side of Eq.~\eqref{eqbolintra}, the $\mathbf B$-dependent currents are given by the terms resulting from $n>0$. In particular, the term with $n=1$ gives us the classical Hall current, which is proportional to linear order in $B$. The $n>1$ terms produce terms depending on higher powers of $B$.
Using the explicit forms of the solutions for $f^\lambda $, the electric- and the thermal-current densities are captured by
\begin{align}
{\mathbf J}
& =   -\, e  \int
\frac{ d^2 \mathbf k}{(2\, \pi)^2 } \,
\boldsymbol{v}^\lambda \,  \delta f^\lambda (\mathbf k)  \text{ and } 
 \mathbf{J}^{\rm th}  =  \int
\frac{ d^2 \mathbf k} {(2\, \pi)^2 } \,
 \boldsymbol{v}^\lambda \left( \epsilon^\lambda - \mu \right)  
 \delta f^\lambda ( \mathbf k)\,,
 \end{align}
respectively. In what follows, we show the results for $\mu>0$ and $ T \ll \mu $, for which the $ \lambda =1 $ band contributes.

\section{Response in the absence of a magnetic field}
\label{seczerob}

In the absence of a magnetic field, the solution for $\delta f^\lambda (\mathbf k)$ is obtained from Eq.~\eqref{eqbolintra} as (see also \cite{mermin, ips-kush-review})
\begin{align}
\delta f^\lambda =  \tau \, f_0^\prime (\epsilon^\lambda ) 
\left [
{\boldsymbol{v}}^\lambda (\mathbf k)
\cdot \left \lbrace  
e \, \mathbf E + \frac{  \epsilon^\lambda (\mathbf k) - \mu } {T}  \, \nabla_{\mathbf r} T 
\right \rbrace \right] .
\end{align}
Consequently, the explicit forms of the electric- and the thermal-current densities [cf. Eq.~\eqref{eqcur1}] are captured by
defining
\begin{align}
\label{eqlgeneric}
& \mathcal{L}^{(\zeta)}_{ij} \equiv 
- \,e^2 \, \tau  \sum_{\lambda=\pm}
\int\frac{d^2\mathbf{k}}{( 2\, \pi )^2}
 \,   f_0^\prime \big (\epsilon^\lambda ({\mathbf k}) \big ) \;
 v^\lambda_i (\mathbf{k}) \, v^\lambda_j(\mathbf{k})
\left [\epsilon^\lambda ({\mathbf k})-\mu \right ]^\zeta \,,
\text{ where } \zeta \in \lbrace 0, 1, 2 \rbrace\,,
\nn \text{such that } \qquad
& \sigma_{ij}^{(0)} =  \mathcal{L}_{ ij }^{(0)} \,, 
\quad
  T\,  \alpha_{ij}^{(0)}
= \frac{- \, \mathcal{L}_{ij}^{(1)}} {e}\,, 
\text{ and }
\ell_{ ij }^{(0)} = \frac{\mathcal{L}_{ij}^{(2)}
}  {e^2 \, T } 
\end{align}
represent the three independent linear-response coefficients arising when the magnetic field is set to zero. We note that, although we have summed over the two bands, either the positive or the negative band contributes for $T \rightarrow 0 $, depending on the sign of $ \mu $.

For our system described in Eq.~\eqref{eqham}, some straightforward calculations (using the results shown in Appendices~\ref{appint} and \ref{appsom}) yield
\begin{align}
\label{eqLvals}
 \mathcal{L}^{(0)}_{xx} & =  
 \frac { - \, e^2 \, \mu \, \tau  \, w_x}
{2 \, \pi   \left (1 - \eta^2 \right)^{\frac{3} {2}}
   \left (\sqrt {1 - \eta^2} - 1 \right)^2  w_y}  \times g_0 \,,\quad
\mathcal{L}^{(1)}_{xx}  =  
\frac {\pi^2 \, T^2} {3\, \mu} \,\mathcal{L}^{(0)}_{xx}  \,,\quad
\mathcal{L}^{(2)}_{xx}  = 
\frac {\pi^2 \, T^2} {3} \,\mathcal{L}^{(0)}_{xx}   \,,
\end{align}
\begin{align}
\label{eqLvals1}
 \mathcal{L}^{(0)}_{xy} = \mathcal{L}^{(0)}_{yx} & =  
  \frac {  e^2 \, \mu \, \tau \, \eta_x \, \eta_y} 
 {2 \, \pi\left (1 - \eta^2 \right)^{ \frac{3}{2}}
 \left (\sqrt {1 - \eta^2} - 1 \right)^2}
  \times g_1 \,,\quad
\mathcal{L}^{(1)}_{xy} = \mathcal{L}^{(1)}_{yx} =  
\frac {\pi^2 \, T^2} {3\, \mu} \,\mathcal{L}^{(0)}_{xy}  \,,\quad
\mathcal{L}^{(2)}_{xy} = \mathcal{L}^{(2)}_{yx}  = 
\frac {\pi^2 \, T^2} {3} \,\mathcal{L}^{(0)}_{xy}   \,,
\end{align}
where
\begin{align}
\label{eqfns}
g_0 & =  
\frac {w_x^2} {\eta^4} \Big [
4 \, \eta^2\left ( 1- \sqrt {1 - \eta^2}  \right)
- \left (\sqrt {1 - \eta^2} - 4 \right)\eta_x^6
+\eta_x^4 \left (
  7 \, \sqrt {1 - \eta^2}
   - \left (\sqrt {1 - \eta^2} - 5 \right) \eta_y^2  - 9 \right)
\nn & \hspace{ 1 cm}
+ \left (
   9 + \left (\sqrt {1 - \eta^2} - 2 \right)\eta^2
    - 7 \, \sqrt {1 - \eta^2}
     \right)\eta_y^4
+ 2\left (\eta^2 + 4 \right)\left (\sqrt {1 - \eta^2} -  1 \right)\eta_y^2
- \eta_y^6
\Big ]
\nn & \hspace{ 0.5 cm}
+
\left (2 - \eta^2 - 2 \, \sqrt {1 - \eta^2} \right)
\left (2 \,  w_x - 1 \right)\eta_x^2 ,\nn
g_1 & = \frac {w_x \, w_y} {\eta^4} \Big [
8\left (\sqrt {1 - \eta^2} - 1 \right)
+4 \, \eta^2 \left (4 - 3\sqrt {1 - \eta^2} \right)
+\left (2 \, \sqrt {1 - \eta^2} - 7 \right)\eta_x^4
+\left (2 \, \sqrt {1 - \eta^2} - 7 \right)\eta_y^2
\left (2 \, \eta^2 - \eta_y^2 \right)
\Big ]
\nn & \hspace{ 0.5 cm}
+ \left ( 2 - \eta^2 - 2\sqrt {1 - \eta^2} \right)
\left ( 1 - w_x - w_y \right).
\end{align}
We would like to point out that the answer for $\mathcal{L}^{(\zeta)}_{yy}$ is easily obtained from that for $\mathcal{L}^{(\zeta)}_{xx}$ (shown above)  simply interchanging the subscripts as
$ \lbrace x , \, y \rbrace \rightarrow  \lbrace y , \, x \rbrace $.
We also find that the Mott relation and the Wiedemann-Franz law are satisfied \cite{mermin} (which are applicable in the $T\rightarrow 0$ limit), due to the relations between (1) $\mathcal{L}^{(0)}_{ij}$ and $\mathcal{L}^{(1)}_{ij}$, and (2) $\mathcal{L}^{(0)}_{ij}$ and $\mathcal{L}^{(2)}_{ij}$, respectively.

For the zero magnetic field case, we have only the components of the currents surviving which are collinear with $\mathbf E  $ and/or $ \nabla_{\mathbf r} T $, as applicable. In particular, for transport along the electric field and the thermal gradient aligned along the same direction, say along the Cartesian axis labelled as $i$, we have
\begin{align}
S_{ii} & = \frac{\alpha_{ii}} { \sigma_{ii} }
\,, \quad
\kappa_{ii}
 = \ell_{ii}- \frac{ T \left( \alpha_{ii} \right)^2 } { \sigma_{ii} }  \,.
\end{align}

\section{Response in the presence of a weak nonquantizing magnetic field}
\label{secb}


The $n \geq1$ terms of Eq.~\eqref{eqbolintra} give rise to the $\mathbf B$-dependent terms, such that
\begin{align}
\label{eqbol2}
&- \,\frac{{\tilde g}^{\lambda} (\mathbf{k}) } 
{  \tau }
=  
\left [   \boldsymbol{v}^\lambda (\mathbf k) \cdot 
\left  \lbrace e\, \mathbf{E} +
\frac{  \epsilon^\lambda (\mathbf k)  - \mu } {T}  
\, \nabla_{\mathbf r} T  \right \rbrace  \right ]
+ \frac{\Upsilon^\lambda (\mathbf k) } 
{\tau} \,, \nn & \text{ where }
\Upsilon^\lambda (\mathbf k) =
\tau   \sum_{n = 1}^{\infty}
\left[ e \, \tau \hat{L} (\mathbf k) \right ]^n 
\left [   \boldsymbol{v}^\lambda (\mathbf k) \cdot 
\left  \lbrace e\, \mathbf{E} +
\frac{  \epsilon^\lambda (\mathbf k)  - \mu } {T}  
\, \nabla_{\mathbf r} T  \right \rbrace  \right ]
=
\tau   \left  \lbrace e\, \mathbf{E} +
\frac{  \epsilon^\lambda (\mathbf k)  - \mu } {T}  
\, \nabla_{\mathbf r} T  \right \rbrace
\cdot
\sum_{n = 1}^{\infty}
\left[ e \, \tau \hat{L} (\mathbf k) \right ]^n 
\,  \boldsymbol{v}^\lambda (\mathbf k)   .
\end{align}
The last identity follows from the fact that $ \hat{L} \, \epsilon^\lambda = 0$. 
We observe that, for $\mathbf B  =\lbrace B_x, \, B_y, \, 0 \rbrace $ (i.e., when the magnetic field is applied in the $xy$-plane), a nonzero $\mathbf B $ results in the same linear response as the $\mathbf B = \mathbf 0 $ case. A nontrivial $\mathbf B $-dependent response is possible only for $\mathbf B = B_z\, {\boldsymbol{\hat e}}_z $.
An alternate way of determining the Lorentz-force contribution for the untilted case
is shown in Appendix~\ref{apploretnz}, which cannot be applied for the tilted cases.

The Lorentz-forced-induced contributions to the electric- and thermal-current densities take the forms of
\begin{align}
\left( J_{\text{\tiny{LF}}} \right)_{i}
& =   -\, e  \sum_{\lambda=\pm} \int
\frac{ d^2 \mathbf k}{(2\, \pi)^2 } \,
{v}^\lambda_i \, \Upsilon^\lambda (\mathbf k)  
\left [- f_{0}^\prime (\epsilon^\lambda (\mathbf k)) \right ] 
\nn \text{and } 
\left(  J^{\rm th}_{\text{\tiny{LF}}}  \right)_i
& = \sum_{\lambda=\pm} \int
\frac{ d^2 \mathbf k} {(2\, \pi)^2 } \,
 {v}^\lambda_i \left [ \epsilon^\lambda (\mathbf k) - \mu \right ]
\, \Upsilon^\lambda (\mathbf k) 
\left [- f_{0}^\prime (\epsilon^\lambda (\mathbf k)) \right ] ,
 \end{align}
respectively. Accordingly, we define
\begin{align}
\label{eqlgeneric}
& \mathcal{L}^{(\zeta), \text{\tiny{LF}}}_{ij} \equiv 
- \,e   \sum_{\lambda=\pm}
\int\frac{d^2\mathbf{k}}{( 2\, \pi )^2}
 \,   f_0^\prime \big (\epsilon^\lambda ({\mathbf k}) \big ) \;
 v^\lambda_i (\mathbf{k}) \, 
 \partial_{E_j} \Upsilon^\lambda_j(\mathbf{k})
\left [\epsilon^\lambda ({\mathbf k})-\mu \right ]^\zeta \,,
\text{ where } \zeta \in \lbrace 0, 1, 2 \rbrace\,,
\nn \text{such that } \qquad
& \sigma_{ij}^{\text{\tiny{LF}}} =  
\mathcal{L}_{ ij }^{(0), \text{\tiny{LF}}} \,, \quad
  T\,  \alpha_{ij}^{\text{\tiny{LF}}}
= \frac{- \, \mathcal{L}_{ij}^{(1), \text{\tiny{LF}}}} {e}\,, 
\text{ and }
\ell_{ ij }^{\text{\tiny{LF}}} = \frac{\mathcal{L}_{ij}^{(2), \text{\tiny{LF}}}
}  {e^2 \, T } \,,
\end{align}
which represent the parts of the three independent linear-response coefficients arising from the nonzero magnetic field.
Alternatively,
\begin{align}
\label{eqlgeneric2}
& \sigma_{ij}^{\text{\tiny{LF}}}  =
- \,e  \sum_{\lambda=\pm}
\int\frac{d^2\mathbf{k}}{( 2\, \pi )^2}
 \,   f_0^\prime \big (\epsilon^\lambda ({\mathbf k}) \big ) \;
 v^\lambda_i (\mathbf{k}) \, \partial_{E_j} \Upsilon^\lambda_j(\mathbf{k})\, ,
\nn & \alpha_{ij}^{\text{\tiny{LF}}}  =
- \,e  \sum_{\lambda=\pm}
\int\frac{d^2\mathbf{k}}{( 2\, \pi )^2}
 \,   f_0^\prime \big (\epsilon^\lambda ({\mathbf k}) \big ) \;
 v^\lambda_i (\mathbf{k}) 
 \left [-  \partial_{\partial_j T} \Upsilon^\lambda_j(\mathbf{k}) \right ],
 \text{ and}
\nn & \ell_{ij}^{\text{\tiny{LF}}}  =
\sum_{\lambda=\pm}
\int\frac{d^2\mathbf{k}}{( 2\, \pi )^2}
 \,   f_0^\prime \big (\epsilon^\lambda ({\mathbf k}) \big ) \;
 v^\lambda_i (\mathbf{k}) 
 \left [-  \partial_{\partial_j T} \Upsilon^\lambda_j(\mathbf{k}) \right ]
\left [ \epsilon^\lambda (\mathbf k) - \mu \right ] .
\end{align}
Note that $\sigma_{ij}^{\text{\tiny{LF}}}$ is nonzero only if $E_j \neq 0$; and $\alpha_{ij}^{\text{\tiny{LF}}}$ and $\ell_{ij}^{\text{\tiny{LF}}} $ are nonzero only if $\partial_j T
\neq 0$. Table~\ref{table_params} shows some typical parameter values for $\alpha$-(BEDT-TTF)$_2$I$_3$.

\subsection{$\mathbf{E}= E_x \,  {\boldsymbol{\hat e}}_x $, $\nabla_{\mathbf r} T =\mathbf 0 $, $\mathbf{B}= B_z
\, {\boldsymbol{\hat e}}_z $}

For the configuration with $\mathbf{E}= E_x \,  {\boldsymbol{\hat e}}_x $, $\nabla_{\mathbf r} T =
\mathbf 0 $, $\mathbf{B}= B_z
\, {\boldsymbol{\hat e}}_z $, we obtain
\begin{align}
\label{equpsex}
\partial_{E_x} {\Upsilon}^{\lambda} (\mathbf{k}) 
& = \frac { \lambda  \,  B_z \, e^2\, \tau^2 \, k_y \, w_x^2 \, w_y^2
\; \epsilon^{\lambda} (\mathbf k) }
{ \varepsilon^3} 
-\frac{
\lambda\, e^3\, \tau^3 \,  B_z^2 \, w_x^3  \, w_y^2} { \varepsilon^5}
\Big[  k_x^3 \,  w_x^3 \left ( 1 +\eta_x^2 \right)
+ 2  \,k_x^2 \, w_x^2 \, \eta_x
\left (2 \, k_y\,  w_y\,\eta_y + \lambda   \,  \varepsilon \right) 
\nn & \hspace{ 0.5 cm } 
+  k_x\,  k_y \, w_x  \,w_y
   \left \lbrace k_y  \,w_y
   \left ( 1 -2 \,\eta_x^2 + 3 \,\eta_y^2  \right) 
 + 4 \, \lambda\,    \varepsilon \,\eta_y \right \rbrace 
 -  2 \, k_y^2\,  w_y^2\, \eta_x
   \left (k_y \, w_y\, \eta_y + \lambda \,  \varepsilon \right) 
 \Big ]
+ \mathcal{O}(B_z^3) \,.
\end{align}
For the sake of brevity, we have shown here the answer correct upto $\mathcal{O}(B_z^2)$.
For the untilted case, we find that
\begin{align}
\label{equntilted}
{\Upsilon}^{\lambda} (\mathbf{k})
= \frac {E_x \, B_z \, e^2\, \tau^2\, w_x^2\, v_y^\lambda }   { \varepsilon} 
\left [ 
\lambda -\frac {\lambda \, B_z \, e \, \tau  \, w_y^2 \, v_x^{\lambda}
    \left (k_x \, v_x^{\lambda} + k_y \, v_y^{\lambda} \right)}
{ \varepsilon^2 \,  v_y^{\lambda}}
-\frac {B_z^2 \, e^2 \, \tau^2 \, w_x^2 \, w_y^2
    \left (k_x \, v_x^{\lambda} + k_y \, v_y^{\lambda} \right)}
{  \varepsilon^3}
\right ] + \mathcal{O}(B_z^4)\,.
\end{align} 
Comparing with Eq.~\eqref{eqcompare} (in Appendix~\ref{apploretnz}), we find that, although the leading-order term matches for both the approaches, they deviate form each other starting from $\mathcal{O} (B_z^2 )$ terms.

\begin{table}[t]
\centering
\begin{tabular}{|c|c|c|}
\hline
Parameter &   SI Units &   Natural Units  \\ \hline
$ w_x $ from Refs.~\cite{goerbig, thesis-fuchs} & $ 3 \times10^{5} $ m~s$^{-1} $ & $0.001$  
\\ \hline
$\tau$ from Ref.~\cite{tauval} & $ 10^{-13} \, \text{s} $ & $152  $ eV$^{-1}$  
 \\ \hline
$\mu$ from Refs.~\cite{goerbig, thesis-fuchs} & $ \sim 1.6\times 10^{-22} $ J 
& $\sim 0.001 $ eV \\ \hline
\end{tabular}
\caption{\label{table_params}The values for the various parameters, used in the plots of conductivity, are tabulated here. While using the natural units, we need to set $\hbar=c=k_{B}=1$.}
\end{table}

\begin{figure*}[t]
\subfigure[]{\includegraphics[width = 0.4 \textwidth]{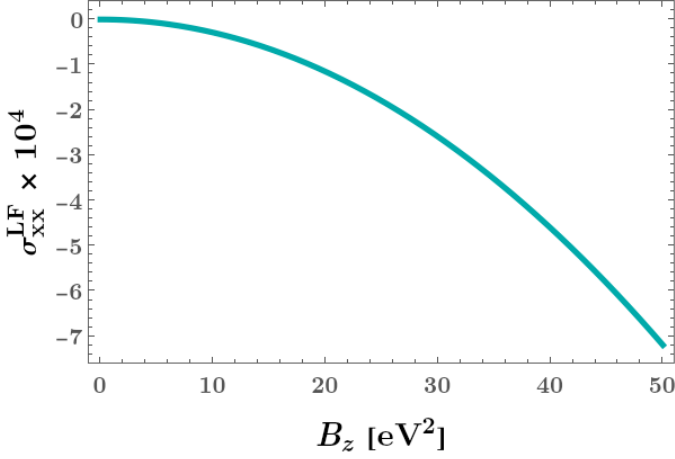}} \hspace{ 1 cm }
\subfigure[]{\includegraphics[width = 0.4 \textwidth]{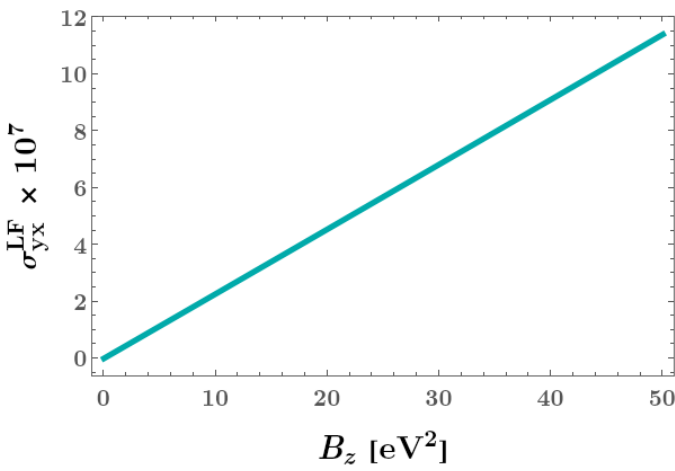}}
\caption{Characteristics of the (a) in-plane longitudinal and (b) out-of-plane components [cf. Eq.~\eqref{eqsiglf}], are shown as functions of $B_z$. We have used $e=1$, $ T = 0 $, $\tau = 152 $ eV$^{-1}$ (see, for example, Ref.~\cite{tauval}), $w_x = 0.001 $, $w_y =  w_x/9$, $\eta_x = 0.34$, $ \eta_y = 0.034$, and $\mu =0.001 $ eV, representing typical parameter values for $\alpha$-(BEDT-TTF)$_2$I$_3$ \cite{goerbig, thesis-fuchs}.
\label{figex}}
\end{figure*}

For $\mu>0$ and $T\ll \mu $ $\Leftrightarrow \lambda =1 $, the nonzero linear-response coefficients are obtained as
\begin{align}
\label{eqsiglf}
\sigma_{xx}^{\text{\tiny{LF}}} 
& = - \,
\frac {e^4 \, \tau^3 \, B_z^2
    \left (3 \, \mu^2 + \pi^2 \, T^2 \right) 
   w_x^4 \, w_y
    \left ( 4 + \eta^2 + 2\, \eta_y^2 \right)}
{48 \, \pi \, \mu^3} + \mathcal{O}(B_z^3) \,,
\nn \sigma_{yx}^{\text{\tiny{LF}}} 
& = 
\frac {e^3 \, \tau^2 \, B_z \, w_x \, w_y^2}
{4\, \pi}
+
 \frac {e^4 \, \tau^3 \, B_z^2
    \left (3 \, \mu^2 + \pi^2 \, T^2 \right) 
   w_x^2 \, w_y^3 \, \eta_x \, \eta_y}
{24 \, \pi \, \mu^3}  + \mathcal{O}(B_z^3)\,.
\end{align}
Here, the $\mathcal{O}(B_z)$ terms represent the classical Hall conductivity due to the Lorentz force and, as expected, it has a vanishing longitudinal part. The higher-order-in-$B_z$ terms represent the terms arising purely due to quantum effects. In Fig.~\ref{figex}, we show two representative curves, for some typical parameter values, as functions of $B_z$. As expected, the $B_z^2$-dependent part is subdominant compared to the linear-in-$B_z$ part in the curve for $\sigma_{yx}^{\text{\tiny{LF}}} $, which justifies our expansion in powers of $B$.

\subsection{$\mathbf{E}= E_y \,  {\boldsymbol{\hat e}}_y $, $\nabla_{\mathbf r} T =\mathbf 0 $, $\mathbf{B}= B_z
\, {\boldsymbol{\hat e}}_z $}

For the configuration with $\mathbf{E}= E_y \,  {\boldsymbol{\hat e}}_y  $, $\nabla_{\mathbf r} T =\mathbf 0 $, and $\mathbf{B} = B_z \, {\boldsymbol{\hat e}}_z $, we obtain
\begin{align}
\label{equpsey}
\partial_{E_y} {\Upsilon}^{\lambda} (\mathbf{k}) 
& = -\, \frac { \lambda  \, B_z \, e^2\, \tau^2 \, k_x \, w_x^2 \, w_y^2
\; \epsilon^{\lambda} (\mathbf k) }
{ \varepsilon^3} + 
\frac { \lambda \, e^3 \, \tau^3 \,  B_z^2 \, w_x^2 \, w_y^3
} { \varepsilon^5}
\Big [
 2\, k_x^3 \, w_x^3 \, \eta_x \, \eta_y
+ k_x^2 \, w_x^2
\left \lbrace k_y \, w_y
\left ( 2 \, \eta_y^2 - 3 \, \eta_x^2 - 1 \right)
   + 2 \, \lambda  \,  \varepsilon \, \eta_y \right \rbrace
\nn & \hspace{ 0.5 cm } 
-4 \, k_x \, k_y \, w_x \, w_y\, \eta_x
\left (k_y \, w_y\, \eta_y + \lambda  \,  \varepsilon \right)
-k_y^2 \, w_y^2 
\left \lbrace  k_y  \, w_y \left ( 1 + \eta_y^2 \right)
+2 \, \lambda \,  \varepsilon \, \eta_y \right  \rbrace
\Big ]
+ \mathcal{O}(B_z^3) \,.
\end{align}

For $\mu>0$ and $ T \ll \mu $ $ \Leftrightarrow \lambda =1 $, the nonzero linear-response coefficients are obtained as
\begin{align}
\sigma_{yy}^{\text{\tiny{LF}}} 
& = -\frac {e^4
   \, \tau^3 \, B_z^2
    \left (3 \, \mu^2 + \pi^2\, T^2 \right) 
   w_x \, w_y^4
    \left ( 4 +  \eta^2 + 2\,\eta_x^2 \right)}
{48 \, \pi \, \mu^3}  + \mathcal{O}(B_z^3)\,,
\nn \sigma_{xy}^{\text{\tiny{LF}}} 
& =
-\frac {e^3 \, \tau^2 \, B_z \, w_x^2\, w_y}
{4\, \pi}
+
\frac {e^4\, \tau^3 \, B_z^2
    \left (3 \, \mu^2 + \pi^2 \, T^2 \right)  w_x^3 \, w_y^2
   \, \eta_x \, \eta_y} {24 \, \pi \, \mu^3}   + \mathcal{O}(B_z^3) \,.
\end{align}
Clearly, the results here are related to the results obtained for the earlier subsection's case by interchanging the subscripts
$ \lbrace x , \, y \rbrace \rightarrow  \lbrace y , \, x \rbrace $ and replacing $B_z \rightarrow -\, B_z$. This is naturally expected from the symmetry of the system.

\subsection{$\mathbf{E}= \mathbf 0 $, $\nabla_{\mathbf r} T = \partial_x T \, {\boldsymbol{\hat e}}_x $, $\mathbf{B} = B_z
\, {\boldsymbol{\hat e}}_z $}

For the configuration with $\mathbf{E}= 
\mathbf 0 $, $\nabla_{\mathbf r} T = 
\partial_x T \, {\boldsymbol{\hat e}}_x $, and $\mathbf{B}= B_z
\, {\boldsymbol{\hat e}}_z $, it is easily observed that 
\begin{align}
 \partial_{\partial_x T} {\Upsilon}^{\lambda} (\mathbf{k}) 
& =
\frac { \lambda \, B_z \, e \,\tau^2  
\,  k_y  \,w_x^2  \, w_y^2}
 {\varepsilon^3 \, T} \,
 \Bigg[
 k_x^2 \, w_x^2 \left ( 1 +\eta_x^2 \right) + 
 k_x  \, w_x \, \eta_x \left (2 \, \varepsilon \, \lambda + 
    2 \, k_y  \, w_y \,\eta_y - \mu  \right)  - \mu
\left ( \varepsilon \, \lambda  + k_y \, w_y \,\eta_y \right) 
\nn & \hspace{ 4 cm }
+ 
 k_y \, w_y\left \lbrace k_y \, w_y\left ( 1 + \eta_y^2 \right) 
  +  2\,\varepsilon \,\lambda \, \eta_y \right \rbrace 
 \Bigg]
\nn & \hspace{ 0.75 cm} -
\frac {e^2 \, \lambda \, \tau^3  \, B_z^2 \, w_x^3  \,w_y^2} 
{\varepsilon^5  \,T}  \,
\Bigg[
k_x^4  \, w_x^4 \, \eta_x \left ( 3 + \eta_x^2  \right)
+k_x^3 \, w_x^3
\left\lbrace
 \varepsilon \, \lambda  + 
 5 \, k_y  \, w_y
 \left ( 1 + \eta_x^2 \right)\eta_y
-\mu
+3 \,\varepsilon \, \lambda \,\eta_x^2 - \mu \, \eta_x^2
\right  \rbrace
\nn & \hspace{ 1 cm}
+k_x^2  \,w_x^2 \, \eta_x
\left \lbrace  k_y  \, w_y
\left (
k_y  \, w_y
\left ( 1-2 \, \eta_x^2 + 7 \, \eta_y^2 \right)
   + 2 \, \eta_y \, (5 \, \varepsilon \, \lambda  - 2 \, \mu) 
   \right)
-2 \, \varepsilon \, \lambda \,  \mu 
\right  \rbrace 
\nn & \hspace{ 1 cm}
+ k_x  \, k_y \, w_x  \, w_y
\left\lbrace
k_y  \, w_y \left ( \varepsilon \, \lambda  - \mu
 + 2 \, \eta_x^2 \, (\mu - 2 \,\varepsilon \, \lambda )
   + \eta_y^2  \, (7 \,\varepsilon \, \lambda  - 3 \, \mu) \right)
+ k_y^2  \, w_y^2 \, \eta_y
\left ( 5-4 \,\eta_x^2 + 3 \,\eta_y^2 \right)
-4 \,\varepsilon \, \lambda  \, \mu \, \eta_y
\right  \rbrace
\nn & \hspace{ 1 cm}
-2  \, k_y^2  \,  w_y^2 \, \eta_x
\left\lbrace
k_y  \, w_y
\left (k_y  \, w_y\left ( 1 + \eta_y^2 \right)
   + \eta_y  (2 \,\varepsilon \, \lambda  - \mu) \right)
- \varepsilon \, \lambda  \, \mu 
\right  \rbrace
\Bigg]
 \,.
\end{align}
In fact, since $\check L \,\epsilon^\lambda = 0 $, we note that
\begin{align}
 \partial_{\partial_x T} {\Upsilon}^{\lambda}  (\mathbf{k}) 
= \partial_{E_x} \Upsilon^{\lambda} (\mathbf{k}) 
\big \vert_{\text{Case A}} \,
\frac {\epsilon^\lambda (\mathbf{k}) - \mu   } {e\, T}  .
\end{align}
Consequently, for$\mu>0$ and $T\ll \mu $ $\Leftrightarrow \lambda =1 $, the nonzero linear-response coefficients are obtained as
\begin{align}
\alpha_{xx}^{\text{\tiny{LF}}}  
& = 
- \,\frac {\pi \, T\, e^3 \, \tau^3 
\, B_z^2  \, w_x^4 \, w_y
\left ( 4 + \eta^2 + 2 \,\eta_y^2  \right)} 
{48 \, \mu^2}
 + \mathcal{O}(B_z^3)\,, \quad
 \ell_{xx}^{\text{\tiny{LF}}}   = \frac{\mu\, \alpha_{xx}^{\text{\tiny{LF}}} 
 } {e}
 + \mathcal{O}(B_z^3)\,,\nn
\alpha_{yx}^{\text{\tiny{LF}}}  
& = 
\frac {\pi \, T\, e^3 \,\tau^3  
\,B_z^2 \, w_x^2\,  w_y^3 \, \eta_x \,\eta_y} 
{24 \, \mu^2}
 + \mathcal{O}(B_z^3)\,, \quad
 \ell_{yx}^{\text{\tiny{LF}}}   = 
 \frac {\pi  \, T\, e \, \tau^2  \, B_z \, w_x \, w_y^2} 
 {12}
+ 
 \frac{\mu\, \alpha_{yx}^{\text{\tiny{LF}}} } 
 {e}
 + \mathcal{O}(B_z^3) \,.
\end{align}
Comparing with Eq.~\eqref{eqsiglf}, we find that the Mott relation and the Wiedemann-Franz law are both satisfied for the $xx$-components. For the $yx$-components, while the Mott relation holds only for the coefficients of the terms proportional to $B_z^2  $, the Wiedemann-Franz law is seen to hold both for the linear-in-$B_z$ and quadratic-in-$B_z$ terms.

\subsection{$\mathbf{E}= \mathbf 0 $, $\nabla_{\mathbf r} T =\partial_y T \, {\boldsymbol{\hat e}}_y $, $\mathbf{B} = B_z
\, {\boldsymbol{\hat e}}_z $}

For the configuration with $\mathbf{E}= \mathbf 0 $, $\nabla_{\mathbf r} T = \partial_y T \, {\boldsymbol{\hat e}}_y $, and $\mathbf{B}= B_z
\, {\boldsymbol{\hat e}}_z $, we note that
\begin{align}
 \partial_{\partial_y T} {\Upsilon}^{\lambda}  (\mathbf{k}) 
= \partial_{E_y} \Upsilon^{\lambda} (\mathbf{k}) 
\big \vert_{\text{Case B}} \,
\frac {\epsilon^\lambda (\mathbf{k}) - \mu   } {e\, T}  .
\end{align}
Hence, the results can be easily obtained from the earlier subsection's case by interchanging the subscripts $ \lbrace x , \, y \rbrace \rightarrow  \lbrace y , \, x \rbrace $ and replacing $B_z \rightarrow -\, B_z$.

\subsection{Generic cases with $\mathbf{B} = B_z\, {\boldsymbol{\hat e}}_z $}

The expressions for linear response for generic cases, with both nonzero $\mathbf E $ and $\nabla_{\mathbf r} T$, each pointing along an arbitrary direction in the $xy$-plane, can be easily deduced from our answers shown in the above subsections.

\section{Summary and outlook}
\label{secsum}

In this paper, we have studied the behaviour of the three independent linear-response coefficients for a tilted 2d Dirac cone with anisotropic Fermi velocities, arising as a consequence of applying spatially- and temporally-uniform probe fields. These probe fields comprise $\mathbf E$ and $\nabla_{\mathbf r} T$, oriented in the plane of the 2d material. Because we have neglected intervalley scatterings, we have considered a single Dirac cone --- the answer for the contributions for multiple Dirac cones in the Brillouin zone can be easily found by adding up the individual contributions. In the future, it will be worthwhile to refine the computations by incorporating the effects of intervalley-collision processes \cite{ips-internode}. After showing the results for the zero-magnetic-field case, we have considered applying a nonzero and uniform $\mathbf B$, such that its magnitude is low-enough to be confined to the nonquantizing regime, where the Landau-level formation can be ignored. Applying the Boltzmann formalism, we have inferred that only when $\mathbf B $ is directed perpendicular to the 2d plane, any nontrivial effect due to the magnetic field arises. The $\mathbf B $-dependent part originates from the so-called Lorentz-force part of the linearized Boltzmann equation, giving rise to electric and thermal currents. While we can correct the answer order-by-order in powers of $B_z$ by a recursive process, the $\mathcal{O}(B_z)$ terms represent the classical Hall-current densities due to the Lorentz force, possessing only the transverse in-plane components as the nonvanishing parts. It is important to note that our recursive expansion, by using integer powers of the Lorentz-force operator ($\check L $), can easily deal with the generic tilted cases. This is to be contrasted with an alternative method used in the literature, which is applicable only for the untilted cones (see, for example, Ref.~\cite{lundgren14_thermoelectric}). In fact, the earlier method fails to capture the correct answer beyond linear-order in the magnetic-field components, as discussed in Appendix~\ref{apploretnz}.

A direction worth exploring is to consider overtilted Dirac cones, when the Fermi surfaces are hyperbolas \cite{marcus-emil, yadav23_magneto, ips-tilted} and, hence, unbounded in the continuum model. Another promising direction is to investigate tilting in semi-Dirac semmimetals, for which the results for the untilted scenarios exist \cite{ips-kush, ips-kush-review}. It also remains to be explored how nontrivial topological properties, such as the Berry curvature and the orbital magnetic moment, emerge when we gap out the Dirac cones \cite{ips-cd, massive-tilted-dirac}, which will then give rise to nonzero in-planar response as well (see Refs.~\cite{ips-rahul-ph, ips-rahul-tilt, ips-ruiz, ips-rsw-ph, ips-shreya, ips-spin1-ph} for some 3d versions). Lastly, an important avenue is to compute the linear response by using many-body techniques like the Kubo formalism \cite{ips-hermann-review}, which can then be easily used to incorporate interactions and disorder \cite{ips-seb, ips_cpge, ips-biref, ips-klaus, rahul-sid, ips-rahul-qbt, ips-qbt-sc, ips-plasmons, ips-jing-plasmons}.

\appendix

\section{Useful integrals}
\label{appint}

For the ease of performing the integrals, we use the coordinate transformation defined by
\begin{align}
\label{eqcoord}
k_x = \frac{ \varepsilon \cos \theta}{  w_x} \,, \quad k_y\ = 
\frac{ \varepsilon \sin \theta} {w_y}\,,
\text{ with }
\varepsilon = \sqrt{ w^2_x \, k^2_x + w^2_y \, k^2_y } \,,
\end{align}
and then the eigenvalues can be rewritten as
\begin{align}
\epsilon^\lambda ({\mathbf k})=  \varepsilon
\left( \eta_x  \cos \theta + \eta_y  \sin \theta 
+ \lambda\right)  \,,
\end{align}
with the Jacobian $ \mathcal J (\varepsilon) = \frac{\varepsilon} { w_x  \, w_y}$.
Using the above, the density-of-states at an energy $ \varepsilon >0$, at $ T = 0 $, turn out to be
\begin{align}
\rho( \varepsilon ) = \int\frac{d^2\mathbf{k}}{( 2\, \pi )^2}
\delta( \varepsilon -\epsilon^+_\mathbf{k}) 
=
\frac{ \varepsilon } { 2\, \pi  \, w_x  \, w_y}
\,    \frac{ 1 } { \sqrt{1-\eta^2} } \,,
\label{Eq_DOS}
\end{align}
where $ \eta = \sqrt{\eta_x^2 + \eta_y^2 } \,.$

\section{Sommerfeld expansion} 
\label{appsom}

In order to perform the integrals involving the derivatives of the Fermi-Dirac distribution function $f_0$, we employ the Sommerfeld expansion \cite{mermin}, which is valid for $ T \ll \mu $. Upon using the Sommerfeld expansion, we get
\begin{align}
& \int_{0}^{ \infty} d \epsilon   \,   
 \epsilon^{n}   \left [ - f^\prime_{0} ( \epsilon )\right ] 
=  {\mathcal F}_n (\mu, T ) \,, \quad
{\mathcal F}_n (\mu, T )=  \mu^{n} \,  \left[ 1 + \frac{\pi^2 \,  n \,  (n-1)}
{6 \left(  \beta \,  \mu \right)^2 } 
+ \frac{7 \, \pi^2 \,n\,(n-1)  \,(n-2) \,(n-3) }
 {360  \left(  \beta \,  \mu \right)^4 } 
+ \order{\left( {\beta \,\mu}\right)^{-6}} \right] ,\nn
&  \int_{0}^{ \infty} d \epsilon  \, \epsilon^n  \,(-1)^{ m +1}  \,
\frac{\partial^{ m+1} 
	\, f_{0} ( \epsilon  ) } { \partial \epsilon ^{ m+1} }  
= \frac{n!}{(n-m)!}  \, {\mathcal F}_{n- m} (\mu, T )\,,\nn
& \int_{0}^{ \infty} d \epsilon   \, \epsilon^{n}  \,  (\epsilon - \mu ) 
\,  (-1)^{ m+1}  
\, \frac{\partial^{ m +1 } f_{0} ( \epsilon ) }  { \partial \epsilon^{ m+1} } 
= \frac{(n+1)!}{(n+1- m )!}  \, 
{\mathcal F}_{n+1- m } (\mu, T )- \mu \, \frac{n!}{(n- m )!}  
\, {\mathcal F}_{n-m} (\mu, T ) \,,
\label{eqsom}
\end{align}
where $\beta =1/T $.

 \section{Lorentz-force part using an alternative method for the untilted case}
 \label{apploretnz}
 
 Let us consider the untilted case with $\boldsymbol{w}_0 = \mathbf 0 $.
 Under the action of a weak nonquantizing magnetic field $\mathbf B$, we need to solve for
\begin{align}
\label{eqbol1}
&   \left [
{\boldsymbol{v}}^\lambda (\mathbf k)
\cdot \left \lbrace  
e \, \mathbf E + \frac{  \epsilon^\lambda (\mathbf k) - \mu } {T}  \, \nabla_{\mathbf r} T 
\right \rbrace \right]
\left [-f_0^\prime(\epsilon^\lambda (\mathbf k)) \right ]
- e \, \check{L} \, \delta f^\lambda (\mathbf k)
 = \frac{  -\,  \delta f^\lambda (\mathbf k)} 
{\tau   } \,,  \text{ where }
\check{L} 
\equiv  \left [ {\boldsymbol{v}}^\lambda (\mathbf k) \cross {\mathbf B}
 \right ] \cdot \nabla_{\mathbf k}
\end{align}
is the Lorentz-force operator.
The solution for $\delta f^\lambda $ is obtained by using the ansatz \cite{sasaki, lundgren14_thermoelectric}
\begin{align}
\label{eqsolapp}
\delta f^\lambda (\mathbf k) =  \tau \, f_0^\prime (\epsilon^\lambda (\mathbf k) ) 
\left [
{\boldsymbol{v}}^\lambda (\mathbf k)
\cdot \left \lbrace  
e \, \mathbf E + \frac{  \epsilon^\lambda (\mathbf k) - \mu } {T}  \, \nabla_{\mathbf r} T 
\right \rbrace \right] 
- f_0^\prime (\epsilon^\lambda (\mathbf k)) \left[
\boldsymbol{v}^\lambda(\mathbf k) \cdot \mathbf \Lambda^\lambda (\mathbf k, \mathbf B, \tau) \right ],
\end{align}
where $ \mathbf \Lambda^\lambda \equiv \lbrace  \Lambda^\lambda_x,  \Lambda^\lambda_y, 0 \rbrace  $ represents the correction term arising due to the magnetic field. Plugging it into Eq.~\eqref{eqbol1} and demanding that the Boltzmann equation holds for all values of $ \boldsymbol v^\lambda $, we determine the components of $\mathbf \Lambda^\lambda$ on a case-by-case basis. Since the action of $\check L $ on the first term of the right-hand side of Eq.~\eqref{eqsolapp} is zero, Eq.~\eqref{eqbol1} leads to the self-consistent condition
\begin{align}
\label{eqlorentz}
&  - e \, \check{L}(\mathbf k) \, \delta f^\lambda (\mathbf k)
 = 
\frac{ f_0^\prime (\epsilon^\lambda (\mathbf k)) \left[
\boldsymbol{v}^\lambda(\mathbf k) \cdot 
\mathbf \Lambda^\lambda (\mathbf k, \mathbf B, \tau) \right ]}
{\tau}
\nn & \Rightarrow 
e \, \check{L} (\mathbf k) \left[ 
 f_0^\prime (\epsilon^\lambda (\mathbf k) ) 
\left \lbrace
{\boldsymbol{v}}^\lambda (\mathbf k)
\cdot \left (  
e \, \mathbf E + \frac{  \epsilon^\lambda (\mathbf k) - \mu } {T}  \, \nabla_{\mathbf r} T 
\right ) 
- \frac{
\boldsymbol{v}^\lambda(\mathbf k) \cdot \mathbf \Lambda^\lambda (\mathbf k, \mathbf B, \tau)
} {\tau } \right \rbrace
\right]  = 
\frac{ -\, f_0^\prime (\epsilon^\lambda (\mathbf k)) \left[
\boldsymbol{v}^\lambda(\mathbf k) \cdot 
\mathbf \Lambda^\lambda (\mathbf k, \mathbf B, \tau) \right ]}
{\tau^2 }
\end{align}

We observe the following trivial cases, when a nonzero $\mathbf B $ results in the same linear response as the $\mathbf B = \mathbf 0 $ case: For $\mathbf B  =\lbrace B_x, \, B_y, \, 0 \rbrace $, i.e., when the magnetic field is applied in the $xy$-plane, $\check L $ evaluates to $ \left( B_y \, v_x^\lambda -B_x\, v_y^\lambda \right) \partial_{k_z} $. Hence, Eq.~\eqref{eqlorentz} reduces to
\begin{align}
e \, \boldsymbol{v}^\lambda  \cdot 
\check{L} \, \mathbf \Lambda^\lambda   = 
 \frac{  \boldsymbol{v}^\lambda  \cdot 
\mathbf \Lambda^\lambda  }
{\tau }
\Rightarrow  
\mathbf \Lambda^\lambda    = \mathbf 0\,.
\end{align}
Therefore, the nontrivial cases correspond to $\mathbf B = B_z
\, {\boldsymbol{\hat e}}_z$, such that $\check L = B_z \left( v_y^\lambda \, \partial_{k_x}
- v_x^\lambda \, \partial_{k_y} \right) $. We also note that $\check L  f_0^\prime (\epsilon^\lambda  ) 
= B_z \left( - v_x^\lambda\, v_y^\lambda  + v_x^\lambda \, v_y^\lambda  \right) f_0^{\prime \prime} (\epsilon^\lambda) = 0 $.

Let us consider the example when $\mathbf{E}= E_x \, {\boldsymbol{\hat e}}_x $, $\nabla_{\mathbf r} T =\mathbf 0 $, $\mathbf{B}= B_z \, {\boldsymbol{\hat e}}_z $. Then, Eq.~\eqref{eqlorentz} reduces to
\begin{align}
& \left( v_y^\lambda \, \partial_{k_x}- v_x^\lambda \, \partial_{k_y} \right)
 \left[ e\, v^\lambda_x \, E_x 
- \frac{
\boldsymbol{v}^\lambda \cdot \mathbf \Lambda^\lambda 
} {\tau } \right]  = 
\frac{ -\,
\boldsymbol{v}^\lambda \cdot  \mathbf \Lambda^\lambda }
{ e\, B_z \,\tau^2 }
 \Rightarrow 
\mathbf \Lambda^\lambda  = 
\frac { E_x \, B_z \, e^2\, \tau^2\, w_x^2}
{  \varepsilon^2  + B_z^2\, e^2\, \tau^2 \, w_x^2 \, 
  w_y^2} 
  \left (B_z \, e\, \tau  \, w_y^2 \, {\boldsymbol{\hat e}}_x
-  \lambda \,  \varepsilon  \, {\boldsymbol{\hat e}}_y \right ),
\end{align}
leading to
\begin{align}
\label{eqcompare}
 - \,\boldsymbol{v}^\lambda \cdot  \mathbf \Lambda^\lambda & = 
-\,\frac {E_x \, B_z  \, e^2 \, \tau^2 \, w_x^2 \, w_y^2 }
{  \varepsilon^2   + B_z^2 \, e^2 \, \tau^2 \, w_x^2 \, w_y^2 } 
\,\frac{ B_z \, e \, \tau \,  v_x^{\lambda} - k_y}
{\varepsilon}
\nn &
= \frac {E_x \, B_z \, e^2\, \tau^2\, w_x^2\, v_y^\lambda }   {\varepsilon} 
\left ( 
\lambda - \frac {B_z \, e \, \tau  \, w_y^2 \, v_x^{\lambda} }
{ \varepsilon \, v_y^{\lambda}} 
-\frac {\lambda \, B_z^2 \,  e^2 \, \tau^2 \, w_x^2 \, w_y^2}
{ \varepsilon^2 }
+ \frac {B_z^3 \, e^3 \, \tau^3 \, w_x^2 \, w_y^4 \, v_x^{\lambda} }
{ \varepsilon^3 \, v_y^{\lambda}}
\right) + \mathcal{O}(B_z^4)\,.
\end{align}
Comparing with Eq.~\eqref{equntilted} in the main text, we find that, although the $ \mathcal{O}(B_z) $ term matches for both the approaches, they deviate from each other starting from the $\mathcal{O} \left(B_z^2\right)$ terms. This is expected, because in the approach discussed here, the solution is obtained by assuming that the Boltzmann equation holds for all values of $ \boldsymbol v^\lambda $, which in general need not be true. Hence, the method of finding the solution using the ansatz in Eq.~\eqref{eqsolapp} and equating the coefficients of $ v^\lambda_x $ and $v^\lambda_y$ for finding the generic terms order-by-order in $B_z$ is problematic. In particular, it does not work even at linear-order in $B_z$ when we have tilt.

\bibliography{ref_qui}

\end{document}